# Uniaxial Néel Vector Control in Perovskite Oxide Thin Films by Anisotropic Strain Engineering

K. Kjærnes[1], I. Hallsteinsen[1,2,#], R. V. Chopdekar[2], M. Moreau[1,4], T. Bolstad[1], I-H. Svenum[5,6], S. M. Selbach[3], T. Tybell[1,*]

[1]Department of Electronic Systems, NTNU - Norwegian University of Science and Technology, Norway
[2]Advanced Light Source, Lawrence Berkeley National Laboratory, Berkeley, California 94720, USA
[3]Department of Materials Science and Engineering, NTNU - Norwegian University of Science and Technology, Norway
[4]Silicon Laboratories Norway AS, Trondheim Office, Norway
[5]SINTEF Industry, Trondheim, Norway
[6]Department of Chemical Engineering, NTNU - Norwegian University of Science and Technology, Norway

[*] *Corresponding author:* thomas.tybell@ntnu.no

## Abstract

Antiferromagnetic thin films typically exhibit a multi-domain state, and control of the antiferromagnetic Néel vector is challenging as antiferromagnetic materials are robust to magnetic perturbations. By relying on anisotropic in-plane strain engineering of epitaxial thin films of the prototypical antiferromagnetic material $LaFeO_3$, uniaxial Néel vector control is demonstrated. Orthorhombic (011)- and (101)-oriented $DyScO_3$, $GdScO_3$ and $NdGaO_3$ substrates are used to engineer different anisotropic in-plane strain states. The anisotropic in-plane strain stabilises structurally monodomain monoclinic $LaFeO_3$ thin films. The uniaxial Néel vector is found along the tensile strained *b* axis, contrary to bulk $LaFeO_3$ having the Néel vector along the shorter *a* axis, and no magnetic domains are found. Hence, anisotropic strain engineering is a viable tool for designing unique functional responses, further enabling antiferromagnetic materials for mesoscopic device technology.

[#] Present address: *Department of Materials Science and Engineering, NTNU - Norwegian University of Science and Technology, Norway*

# Introduction

Antiferromagnetic (AF) materials have over the past decade taken important steps to become the active ingredient in novel energy-efficient spintronic devices [1-4]. AF materials are intrinsically spin compensated on the unit cell level, supporting a coherent spin structure essentially free of stray magnetic fields down to very small volumes. Moreover, AF materials display ultrafast dynamic responses, are non-volatile, and can sustain spin-currents over several micrometres [2, 3, 5]. Hence, AF materials possess several of the key features necessary for dense packing and ultrafast dynamics in memory applications. On the downside, however, AFs are exceedingly difficult to control by external magnetic fields, there is no straightforward way to control them by voltage, and they easily form magnetic domains [1]. Therefore, in commercial devices thus far, AFs have mainly been used as a passive pinning-layer for an adjacent ferromagnetic (FM) material via exchange bias.

In order for AF materials to take the active role, methods need to be found for engineering AFs with proper domain control supporting uniaxial spin structures at relevant length scales, which can be switched between stable states. With regards to switching, recent studies have shown that in metallic AFs it is possible to control AF domains using electric currents [6-10]. However, current-driven devices suffer from resistive Joule heating, and so insulating AFs could potentially enable spin transport without charge transport in novel energy-efficient spintronic devices [5, 11-13]. In terms of AF domains, AF materials generally display domain formation, which affects spin transport [14]. To minimise domains with an unfavourable spin axis orientation thus remains an issue for taking full control of the spin texture and Néel vector over macroscopic as well as microscopic regions [15-17]. Strain engineering is an interesting tool in this regard, both for enhancement and tuning of functional properties [18]. It was recently predicted by Density Functional Theory (DFT) calculations, and demonstrated by use of piezoelectric strain, that strain mediated anisotropy can be used to control the Néel vector in manganese-based intermetallic AF systems [19, 20]. Also, structurally twin-free LaFeO$_3$ (LFO) films on GdScO$_3$ (GSO) were recently reported [21]. Nevertheless, the lack of AF domain control still remains to be resolved.

The focus of this work is on anisotropic strain engineering as tool to control the spin structure through magnetocrystalline anisotropy. To this end, the prototypical AF insulator LFO is investigated. LFO exhibits the G-type AF structure with spin-polarised (111)-planes and a high Néel temperature, $T_\text{N} \sim 740$ K, in bulk, the highest of all the orthoferrites [22-27]. The diamagnetic nature of La$^{3+}$ ensures magnetic stability for temperatures below $T_\text{N}$, and the orthorhombic structure (space group 62, $Pbnm$) provides structural stability up to above 1200 K [28, 29]. However, crystal twinning is prevalent [30] due to LFO being ferroelastic [31, 32], which makes it common to observe AF domains [33-36]. To take control of the AF domain structure, anisotropic strain engineering with highly distorted orthorhombic substrates is utilised for growth of monodomain LFO thin films. Two variants of the pseudo-cubic (111)-facet found in orthorhombic systems are used to engineer different anisotropic in-plane strain states. In the cases where the in-plane strain anisotropy is large as experienced by LFO, structurally monodomain thin films emerge where the AF structure also turns monodomain with a uniaxial Néel vector. These findings open for engineering of AF systems with domain control and spin-axis orientation by design.

## Methods

In order to investigate how anisotropic stain affects antiferromagnetism in LFO, epitaxial thin films were deposited on substrates of cubic and orthorhombic symmetry by pulsed laser deposition (PLD). The substrate temperature was 540°C and the substrate-target distance 45 mm for all samples, with a heater temperature ramp rate of 15°Cmin$^{-1}$ for both heating and cooling. An oxygen background of 0.3 mbar was used during heating and deposition, and 100 mbar during cooling. A KrF excimer laser ($\lambda = 248$ nm) with a fluence of $\approx$ 2 Jcm$^{-2}$ and 1-3 Hz repetition rate was used to ablate material from a sintered stoichiometric LFO target [37]. A series of pseudo-cubic (111)-oriented substrates were utilised; DyScO$_3$ (DSO), GdScO$_3$ (GSO) and NdGaO$_3$ (NGO), all orthorhombic (space group 62, $Pbnm$) [38, 39] and isostructural to bulk LFO [30], and cubic SrTiO$_3$ (STO) (space group 221, $Pm\bar{3}m$) [40]. The orthorhombic materials all possess the Glazer tilt pattern [41] $a^-a^-c^+$, and the distortion from ideal cubic perovskite is large in GSO and DSO, intermediate in NGO, and essentially zero in STO [38]. All substrates are commercially available and were received with standard surface miscut, typically ±0.3°, from Shinkosha (Japan), SurfaceNet (Germany), and PI-KEM (England). Two orthorhombic substrate facets were used, (101)$_o$ and (011)$_o$, corresponding to two distinct variants of the (111)$_{pc}$ facet, as illustrated in Figure 1. The notation used to index a symmetry is from here on subscript c for cubic, r for rhombohedral, o for orthorhombic, m for monoclinic, t for triclinic, and pc for pseudo-cubic. The pseudo-cubic unit cell is defined in accordance with *Pbnm* notation, i.e. [001]$_{pc}$ // [001]$_o$ and [110]$_{pc}$ // [100]$_o$, resulting in (101)$_o$ and (011)$_o$ corresponding to $(111)_{pc}$ and $(\bar{1}11)_{pc}$, respectively. The upper row of Figure 1 illustrates the response of the buckled hexagon of the $(111)_{pc}$ surface to the strain anisotropy of the two different orthorhombic facets. The two facets promote an opposite deformation relative to the pseudo-cubic axes, where $(101)_o$ and $(011)_o$ gives relative elongation along $x = \langle 1\bar{1}0 \rangle_{pc}$ and $y = \langle 11\bar{2} \rangle_{pc}$, respectively. The lower row shows the two facets and how the pseudo-cubic unit cell relates to the orthorhombic unit cell with four formula units, i.e. $V_o \approx (\sqrt{2} \times \sqrt{2} \times 2)a_{pc}^3$, needed to correctly describe the distortions and octahedral rotations. The two facets are found by tilting the orthorhombic unit cell around the *a* or *b* axes, respectfully, as illustrated.

A visual representation of the strain conditions investigated is shown in Figure 2. The parameter $\zeta_{ab} = \frac{b_o - a_o}{a_o}$ is introduced as a measure of the degree of orthorhombic distortion in terms of relative difference between $a_o$ and $b_o$ axes. For cubic STO, it follows that $\zeta_{ab}$ is zero. The average pseudo-cubic strain is defined as $\varepsilon_{pc} = \frac{a_{pc(sub)} - a_{pc(film)}}{a_{pc(film)}}$, where $a_{pc} = \frac{\sqrt{a_o^2 + b_o^2 + \frac{c_o}{2}}}{3}$ is the pseudo-cubic parameter averaged over all three components of the orthorhombic unit cell. The anisotropic in-plane strain is defined by $\varepsilon_{x,y} = \frac{l_{x,y(sub)} - l_{x,y(film)}}{l_{x,y(film)}}$, where $l_x$ and $l_y$ is the length of the substrate or film unit cell along the $\langle 1\bar{1}0 \rangle_{pc}$ and $\langle 11\bar{2} \rangle_{pc}$ directions, respectively (see Figure 1). The series of substrates give access to an average strain range from $\varepsilon_{pc} = -1.72\ \%$ compressive strain in NGO to $\varepsilon_{pc} = +1.02\ \%$ tensile strain in GSO, for epitaxial LFO thin films. The strain depends strongly on in-plane crystal direction and substantially varies from the average strain. For example, for LFO on DSO$(101)_o$, along $[010]_o$ a tensile strain of 2.94 % is expected, whereas along $[10\bar{1}]_o$ a compressive strain of 0.22 % is expected. Moreover, it should be noted that for a given substrate material, the two pseudo-cubic (111) facets, $(011)_o$ and $(101)_o$, correspond to different in-plane strain values albeit they have the same orthorhombic distortion. This is particularly well demonstrated for LFO on DSO and GSO, where the (011)$_o$ facet offers both substantial compression and elongation along the two principal in-plane axes, while the (101)$_o$ facet enables a strong elongation along one direction and almost no strain along the other.

All substrates were surface-treated prior to deposition to ensure step-and-terrace surface quality. A standard cleaning routine of 5 min + 5 min of ultrasonication in acetone and ethanol followed by drying in N$_2$ flow and annealing at 1000-1050°C for 1-6 hours was used for all substrates. Annealing was done in a closed furnace with pure O$_2$ flow, heating and cooling with 5°C/min. NGO(011)$_o$ and STO(111)$_c$ were additionally etched in commercial grade hydrofluoric acid buffered in ammonium fluoride (NH$_4$F:HF = 7:1) for 30-60 s directly after cleaning and before annealing. *In-situ* reflection high-energy electron diffraction (RHEED) data is consistent with initial 2D layer-by-layer growth for deposition on all substrates [42]. Atomic force microscopy (AFM) confirmed step-and-terrace topography of the grown samples with RMS roughness typically less than 500 pm.

Four circle high resolution x-ray diffraction (XRD, Bruker D8 Discover) with Cu K-$\alpha_1$ radiation was used to characterise the crystalline quality and symmetry of the thin films. Approximately 17 nm thick films were used in this study to allow for a sufficiently large signal to noise ratio to determine the film symmetry by XRD. Rocking curves from the symmetric (111)$_{pc}$ reflections were used to assess growth quality, and full width at half maximum values were between 0.025° and 0.031° for all samples, of the same order as the substrates and consistent with high quality thin films. The film thickness was confirmed by fitting the $\theta/2\theta$ data, and all samples exhibited clear thickness fringes as shown in Figure 3, indicating coherent growth. The data is plotted in order of decreasing substrate (111)$_{pc}$ lattice constant (increasing $2\theta$), and the two vertical lines indicate the bulk value for the (011)$_o$ and (101)$_o$ LFO facets. It is noted that the NGO(101)$_o$ is not phase pure as the substrate shows a significant portion of (011)$_o$ domains. Also, LFO on NGO(011)$_o$ shows less thickness fringes and lower intensity, indicative of strain relaxation and domain formation. To determine crystal structure and strain state of the thin films, reciprocal space maps (RSM) with linear scans in momentum space ($Q_z = Q_\perp, Q_{x,y} = Q_\parallel$) were used. The RSM data were collected from asymmetric reflections corresponding to pseudo-cubic (312), (132), (330), and (114) reflections as well as symmetric (222) reflections for all samples. Both grazing exit (+) and grazing incidence (-) geometries were utilised in order to account for limited signal intensity due to varying structure factors for some reflections used. It is noted that grazing incidence geometry results in a larger Bragg signal from the thin film at the expense of additional diffuse scattering. The RSM data are depicted in Figure 6 and will be thoroughly presented and discussed later.

In order to corroborate the obtained crystal symmetries, the crystal structure of LFO under strain was also investigated by Density Functional Theory (DFT) calculations. The calculations were done with the Vienna *Ab initio* Simulation Package (VASP) [43, 44] employing the projector augmented wave method (PAW) [44, 45]. The La, Fe and O pseudopotentials supplied with VASP were used, treating 11, 14 and 6 electrons as valence states, with plane-waves expanded up to a cutoff energy of 550 eV. Exchange correlation was described with the Perdew-Burke-Ernzerhof generalised gradient approximation for solids (PBEsol) [46]. A Hubbard U correction following the Dudarev approach [47] was applied. For the Fe 3d states U = 3 eV was used [48]. For La a large U-value is required to minimise f-electrons in the conduction band [49], here U = 8 eV was used, the largest U-value maintaining $a_o < b_o$ for bulk LFO. To calculate the crystal structure for strain parallel to the (111)$_{pc}$ facets, the supercells are reorientations of the cubic or orthorhombic cells into $2(\sqrt{2} \times \sqrt{2} \times \sqrt{3})_{pc}$ pseudo-hexagonal cells with 120 atoms. The in-plane *a* and *b* vectors of the supercells were fixed to be equal to the DFT-calculated bulk values of the substrate materials, while the out-of-plane *c* vector was allowed to relax [50]. Brillouin zone integration was done on a Γ-centred 3x3x3 k-point mesh for the strained LFO supercells. The geometry was optimised until all forces on the ions were below 0.001 eVÅ$^{-1}$. Each orthorhombic facet was initialised with four different settings for the LFO supercell; 1) $a^-a^-c^+$ with the atomic coordinates from bulk LFO, 2) $a^-a^-c^0$ with the atomic coordinates from bulk LFO and zero long-axis octahedral rotations, 3) $a^-a^-a^-$ with atomic coordinates from a rhombohedral unit cell, and 4) $a^-a^-c^+$ with atomic coordinates from the more distorted substrate unit cell of the corresponding facet. These initialisations were chosen to explore possible LFO ground states under anisotropic in-plane strain. All calculations were initialised with collinear G-type antiferromagnetic order on the Fe sublattice of LFO.

To assess the AF Néel vector orientation of the strained LFO films, x-ray absorption spectroscopy (XAS) was performed on beamline 4.0.2 at the Advanced Light Source (ALS), Lawrence Berkeley National Lab, in total-electron-yield mode by monitoring the sample drain current. Linearly polarised light was utilised to probe the x-ray magnetic linear dichroism (XMLD), with the XMLD defined as the difference signal between p- and s-polarised x-rays. XAS spectra were collected from the $L_2$ and $L_3$ edges of Fe and averaged over 4-6 subsequent scans per polarisation, and the magnetic spin axis is interpreted in particular based on the detailed features of the $L_{2A}$ and $L_{2B}$ multiplet. The $L_{2A}$ peaks for p- and s-polarisation were normalised and the relative magnitude of $L_{2B}$ with respect to $L_{2A}$ was used for interpretation of the Néel vector alignment [51-54]. Figure 4a displays typical Fe $L_2$ and $L_3$ edge spectra and the dichroism signal, and the inserts show the L$_2$ edge before (b) and after (c) normalisation of the L$_2$ edge multiplet. Normalisation is done by adjusting the L$_2$ pre-edge to zero and the L$_{2A}$ edge to one. A polarisation signal with relative L$_{2B}$ magnitude above 1.4 is considered perpendicular to the spin axis, and a polarisation signal with relative L$_{2B}$ magnitude below 0.7 showing a double peak feature at the L$_{2B}$ high energy tail is considered parallel to the spin axis [51-54]. The primary measurement geometry is shown in Figure 5, where the incoming x-rays have the in-plane projection along $[11\bar{2}]_{pc}$. In this geometry, s-polarisation probes in-plane along $[1\bar{1}0]_{pc}$ and p-polarisation probes out-of-plane or in-plane along $[11\bar{2}]_{pc}$ depending on the incidence angle $\theta$. The incidence angle was varied as $\theta \in [15°, 165°]$ with reference to the surface plane, where particularly $\theta = 35°, 90°, 145°$ were used for all samples. This is illustrated in the bottom part of Figure 5. At $\theta = 35°$, the photons are incident parallel to the substrate *c* axis, hence p- and s-polarisation probe the orthorhombic *a* and *b* axes (*b* and *a* axes) for a thin film grown on the (101)$_o$ facet ((011)$_o$ facet), respectively. These axes represent the $[110]_{pc}$ and $[1\bar{1}0]_{pc}$ in the *ab*-plane. Probing along two axes belonging to the same (hkl)-family minimises the influence of crystal field anisotropy effects on the XMLD spectra [51, 53]. Moreover, by varying the incidence angle, the p-polarisation gives good coverage of a Néel vector component in the plane of incidence, without having to rotate the sample.

Furthermore, in order to investigate possible AF domain structure, x-ray photoemission electron microscopy (XPEEM) with linearly polarised light was done on the PEEM3 endstation, beamline 11.0.1 at ALS. The XPEEM microscope has a fixed x-ray incidence angle of $\theta = 30°$ with reference to the sample surface, and the incident light was projected along the $[11\bar{2}]_{pc}$ direction with either s- or p-polarised light, similar to the XAS setup as seen in Figure 5. The XMLD contrast at the Fe $L_2$ edge is calculated as defined for the XAS in a pixelwise manner in the field of view, so as to reveal contrast between areas which are more or less magnetically aligned with the incident x-ray **E** vector.

## Results and discussion

First the effect of anisotropic strain on the crystal structure of LFO is assessed. The RSM data for LFO on DSO(101)$_o$, DSO(011)$_o$, GSO(101)$_o$, NGO(101)$_o$, NGO(011)$_o$, and STO(111)$_c$ are given in Figure 6. The calculated unit cell parameters and refined crystal symmetries are tabulated in Table 1, and all RSM data are plotted 1:1, i.e. the in-plane and out-of-plane components are to scale for all plots. Also note the dashed horizontal line in each plot, denoted the symmetry line, which corresponds to the out-of-plane value $Q_\perp$ for a system where all planes measured are symmetric to its mirrored twin, i.e. a high-symmetry cubic or rhombohedral system.

In Figure 6a RSM data for LFO on DSO(101)$_o$ is plotted. All LFO reflections have the same in-plane component $Q_\parallel$ as the substrate, consistent with coherent growth along both principal in-plane axes $[010]_o$ and $[10\bar{1}]_o$. The $(424)_o$ and $(4\bar{2}4)_o$ planes additionally show a broadened, weak signal at higher $Q_\parallel$, indicating an onset of partial relaxation. This can be rationalised due to a large tensile strain of 2.94 % along this direction. The peak positions are symmetric for $(424)_o$ and $(4\bar{2}4)_o$ indicating no unit cell tilt along the $[010]_o$ ($[1\bar{1}0]_{pc}$) axis. From the $(208)_o$ and $(600)_o$ reflections it is apparent that the film is slightly tilted along the $[10\bar{1}]_o$ ($[11\bar{2}]_{pc}$) axis, seen as a collective lowering in the out-of-plane components $Q_\perp$ with respect to the symmetry line. An asymmetry is expected in the $Q_\perp$ values for the $(208)_o$ and $(600)_o$ reflections due to the bulk orthorhombic unit cell both for substrate and film. In order for the strained LFO to be orthorhombic, the LFO $Q_\perp$ values would need to be identical to the substrate values, which is not the case. Thus, a reduction of symmetry to a monoclinic unit cell for LFO on DSO(101)$_o$ can be concluded, see Table 1 for the refined details. DFT calculations for LFO on DSO(101)$_o$ are consistent with the XRD data, showing a monoclinic unit cell 13.9 meV/f.u. above the bulk LFO ground state, with a corresponding Glazer tilt pattern of $a^-a^-c^+$ and space group 14 *P21/n*.

On DSO(011)$_o$, see Figure 6b, the results are similar to DSO(101)$_o$, but with a difference for the $(028)_o$ and $(060)_o$ reflections, where the out-of-plane $Q_\perp$ values for LFO are collectively increased from the symmetry line instead of decreased. The film peak is hidden by the substrate peak in the case for $(028)_o$, but by considering the diffuse scattering around the substrate peak a film peak at slightly lower $Q_\perp$ than for $(060)_o$ is inferred. Hence, there is a small asymmetry in the opposing planes along the $[0\bar{1}1]_o$ ($[1\bar{1}2]_{pc}$) axis, and as DSO is orthorhombic, the in-plane anisotropy due to the epitaxial strain induces a lowering of LFO symmetry to monoclinic. DFT calculations for LFO on DSO(011)$_o$ results in a monoclinic state $a^-a^-c^+$ (14, *P21/b*), 34.9 meV/f.u. above bulk LFO, which corresponds well with the RSM data.

The situation for LFO on GSO(101)$_o$ is comparable to DSO(101)$_o$, see Figure 6c. The crystal symmetry is found to be monoclinic, but there is more partial relaxation than for LFO on DSO(101)$_o$ along the in-plane $[010]_o$ ($[1\bar{1}0]_{pc}$) axis, in agreement with an increased tensile strain to 3.11 %. Note that for the $(424)_o$ and $(4\bar{2}4)_o$ scans, the film reflections are partly hidden by the substrate, however, the film features were resolved using grazing incidence scans to increase film signal and diffuse contributions from nearby regions. The strained $Q_\perp$ values are inferred from relaxation lines following the diffuse scattering. Albeit partial relaxation is readily visible, RSM data for $(532)_o/(5\bar{3}2)_o$ and $(512)_o/(5\bar{1}2)_o$ (not shown) confirm the strained film along the in-plane $[010]_o$ ($[1\bar{1}0]_{pc}$) axis as well. The DFT data for LFO on GSO(101)$_o$ is similar to what was calculated for DSO(101)$_o$. In concordance with the RSM data, the DFT calculations point towards a ground state $a^-a^-c^+$ (14, *P21/n*), which is 21.7 meV/f.u. above the bulk ground state of LFO.

Next, the crystal symmetry of LFO on NGO is considered. LFO on NGO(011)$_o$, see Figure 6d, shows a less coherent and more relaxed film than any of the other cases. The $(244)_o$ and $(\bar{2}44)_o$ scans are symmetric both in and out of plane, and the $(028)_o$ and $(060)_o$ are asymmetric both in and out of plane. The XRD data thus indicates structural domains present in the LFO film, and the RSM signals are spatially averaging many of these. In Figure 7 the three structural domains an orthorhombic film can take on a (111)-oriented monodomain surface are shown. Each domain, A, B, C, is distinguished by the long-axis with in-phase octahedral rotations, which can be distributed among all three pseudo-cubic axes. In addition, each domain may occur with their 90° base-plane rotated twins, A', B', C', where the $a_o/b_o$ axes are swapped, resulting in up to six possible structural domains in total. However, upon increasing the distortion of the unit cell from ideal cubic through rhombohedral to orthorhombic, the likelihood of crystal twinning and structural domains is reduced. Returning to the LFO on NGO(101)$_o$ data, it thus appears that LFO is relaxed to the extent where it forms two or more structural domains. Based on the peak positions of the data, however, the symmetry of LFO reduces to monoclinic also in this case, and the average crystal structure is calculated based on these positions, see Table 1. DFT calculations for LFO on NGO(011)$_o$ finds a ground state $a^-a^-c^{+(-\delta)}$ (14, *P21/b*) at 56.4 meV/f.u. above bulk LFO, where a minor out-of-phase rotation (-δ) is superimposed on the ordinary in-phase rotations along the long axis, i.e. $c^+ \rightarrow c^{+(-\delta)}$.

For LFO on NGO(101)$_o$, Figure 6e, the picture is more complicated than for (101)$_o$ DSO and GSO. The film peaks are broadened in $Q_\parallel$ indicating domains, and some degree of relaxation towards bulk LFO values can be seen. In this case, the $\theta/2\theta$ scan for LFO on NGO(101)$_o$ (Figure 3) reveal a significant contribution from (011)$_o$ domains in the substrate, and the $(424)_o$ and $(4\bar{2}4)_o$ scans show explicitly the signature of another domain in the substrate (Figure 6e red arrows/labels). This signature corresponds to (011)$_o$ domains oriented as $a^+b^-b^-$ (A' domains, Figure 7) with respect to the primary $a^-a^-c^+$ (101)$_o$ (C domain, Figure 7). Thus, the additional substrate peaks seen in the RSM data correspond to reflections of $(152)_o$ and $(\bar{1}36)_o$, where the in-plane components are similar whereas the out-of-plane components differ substantially, as compared to $(424)_o$. The (011)$_o$ domains are in this case oriented with the $[11\bar{1}]_o$ direction parallel to the main $[010]_o$ ($[1\bar{1}0]_{pc}$) axis. As for LFO on NGO(011)$_o$, the film peak positions are used for calculation of an average unit cell, since the film thickness is not sufficient for the RSM data to resolve individual film domains. Interestingly, for LFO on NGO(101)$_o$, the RSM data shows an asymmetry along in-plane $[010]_o$ ($[1\bar{1}0]_{pc}$) rather than $[10\bar{1}]_o$ ($[11\bar{2}]_{pc}$), seen as a splitting of out-of-plane $Q_\perp$ values for LFO peaks in $(424)_o$ and $(4\bar{2}4)_o$ scans relative to the symmetry line. The $(208)_o$ and $(600)_o$ LFO reflections instead fall on similar $Q_\perp$ values. This pattern is consistent with a lowering of LFO symmetry to triclinic due to the unit cell being tilted slightly towards $[010]_o$. Details for the unit cell refinement can be found in Table 1. The DFT results for LFO on NGO(101)$_o$ correspond fairly well with the RSM data and gives the ground state $a^-a^-c^+$ (14, *P21/n*) at 25.7 meV/f.u. above the bulk ground state.

Lastly, LFO on STO(111)$_c$ is included as a cubic system for reference where the substrate strain is isotropic. As seen in Figure 6f, all reflections have the same in-plane component $Q_{\parallel}$ as the substrate, consistent with coherent growth and fully strained films along both principal in-plane axes $[1\bar{1}0]_{pc}$ and $[11\bar{2}]_{pc}$. The peak positions of $Q_{\perp}$ for all film reflections are constant, as is the case for the substrate reflections, and consistent with a compressively strained film. This implies no unit cell tilt along either of the principal in-plane axes $[1\bar{1}0]_{pc}$ and $[11\bar{2}]_{pc}$. Hence, the RSM data is consistent with a rhombohedral unit cell, with a reduced rhombohedral angle of $\alpha_r = 59.34°$ from the ideal cubic value $\alpha_{r,STO} = 60°$, in accordance with an out-of-plane elongation in real space due to the compressive strain. This effectively yields an increase of symmetry in comparison to bulk LFO, indicating a polydomain film. However, DFT calculations for LFO on STO(111)$_c$ reveal a similar monoclinic structure as for the orthorhombic facets, with a $a^-a^-c^+$ (14, *P21/n*) state in agreement with a (101) pseudo-orthorhombic unit cell. This discrepancy between the monoclinic unit cell found by DFT calculations and rhombohedral by XRD can be explained by taking into account structural domain formation. LFO on STO(111)$_c$ is reported to exhibit three to six structural twins (Figure 7) due to the symmetric biaxial strain [36, 55, 56], corresponding to a threefold rotation of the monoclinic unit cell found by DFT calculations. The x-ray beam in the present XRD study has a macroscopic footprint, about $0.4$ mm $\times 10$ mm out from the source monochromator, and the film thicknesses here do not allow for separation of different domains, effectively making the diffracted signal a spatial average over multiple structural domains. However, by disregarding the octahedral rotations in the DFT results, i.e. neglecting the oxygen atoms which have the lowest scattering cross section and considering only the cation positions, the structure obtained from DFT calculations indeed has rotation symmetry around the (111)$_{pc}$ normal. The unit cell is not tilted along either of the in-plane axes due to the symmetric STO strain, and it is thus in agreement with the rhombohedral signature found by XRD.

Taken together, the DFT and RSM results agree well for the crystal structure of LFO thin films on all pseudo-cubic (111) substrate facets included, all data summarised in Table 1. DFT calculations with initialisations based on $a^-a^-c^+$ and $a^-a^-c^0$ tilt patterns give similar results for each facet, whereas a $a^-a^-a^-$ initialisation gives the same crystal symmetry but different space group. The $a^-a^-c^+$ initialisation always results in the lowest energy for the strained state. Based on the crystal structures calculated by DFT, the unit cell of LFO clearly adopts the in-plane anisotropy of the orthorhombic substrates, see Figure 8 where the ground state DFT structures ($a^-a^-c^+$ LFO) are visualised. The monoclinic unit cell parameters are plotted in the upper two panels, whereas the two lower panels show the Fe-O-Fe buckling angles and Fe-O bond lengths along pseudo-cubic axes. All panels are plotted versus in-plane strain along the $x = \langle 1\bar{1}0 \rangle_{\text{pc}}$ direction. The vertical dashed lines denote which data correspond to LFO on each facet referenced on top. The blue dashed line is the linear strain line corresponding to the in-plane $a_o$ and $b_o$ axes for $(011)_o$ and $(101)_o$ facets, respectively. Yellow dashed lines are guides to the eye showing the expected response for the out-of-plane axes. It is noted that the experimental RSM data follow the same trends, only with different absolute strain values due to the difference between DFT calculated and experimental lattice parameters. When the substrate distortion is large, as in the scandates (see Figure 2), the in-phase octahedral rotations increase in amplitude, yielding a reduction of Fe-O-Fe buckling angles along $a_{\text{pc}}/b_{\text{pc}}$. On the other hand, for LFO on NGO where the substrate distortion is smaller than for the scandates (see Figure 2), the in-phase rotations decrease in amplitude, and the buckling angles along $a_{\text{pc}}/b_{\text{pc}}$ increases. The out-of-phase rotations, represented by the bond angles along $c_{\text{pc}}$, seem to depend mostly on facet, as all (011)$_o$ facets yield increased angles, whereas all (101)$_o$ facets give decreased bond angles and hence more buckling. The latter can be directly linked to the deformation settings seen in Figure 1 and Figure 2, where relative elongation and compression of LFO occurs along the in-plane projection of the $c_o$ axis for (011)$_o$ and (101)$_o$, respectively. Following the same trend, all (011)$_o$ facets have longer Fe-O bonds along the $c_{pc}$ axis, and all (101)$_o$ facets have shorter $c_{pc}$ axis bonds. In addition, NGO now has all Fe-O bonds along pseudo-cubic axes shorter than in bulk LFO, whereas the highly distorted scandates split $a_{pc}/b_{pc}$ and $c_{pc}$ bond lengths above and below bulk LFO values depending on facet. Importantly, the XRD RSM data indicate structurally monodomain samples when grown on the highly distorted orthorhombic substrates.

XMLD data

The effect of anisotropic strain on the antiferromagnetic structure of LFO was investigated by x-ray absorption spectroscopy. For grazing incidence of $\theta = 35°$, p-polarised x-rays have the ***E*** vector aligned with the bulk LFO Néel vector along the *a* axis $[100]_o$, about 55° out-of-plane from the (111)$_{pc}$ surface, or the alternative in-plane *b* axis $[010]_o$ for s-polarised x-rays. Based on previous studies, it is expected to find the Néel vector along one of these axes [33, 54, 57], i.e. the $[110]_{pc}$ or $[\bar{1}10]_{pc}$ axes, respectively. Figure 9 shows the total electron yield signal for s- and p-polarisation at incidence angles $\theta = 35°, 90°, 145°$ for all samples. In the following $\theta = 90°$ will correspond to normal incidence, and $\theta = 35°, 145°$ are denoted opposite grazing incidences. In this geometry the spectra are measured with the plane of incidence parallel to the in-plane axes $[10\bar{1}]_o$, $[01\bar{1}]_o$, and $[11\bar{2}]_c$ for facets (101)$_o$, (011)$_o$, and (111)$_c$, respectively. For each sample, the p-polarisation signal will change if the x-ray ***E*** vector becomes more or less aligned with the Néel vector as a function of incidence angle $\theta$, while the s-polarisation signal should be constant with $\theta$, possibly varying slightly due to the footprint being dependent on incidence angle. The relative magnitudes of the $L_{2A/B}$ peaks are thus decisive for the interpretation of the Néel vector alignment.

First the XMLD analysis of LFO with structural domains is presented. As previously reported, LFO on STO(111)$_c$ has a symmetric in-plane response with multiple AF domains aligned along the six in-plane $\langle 1\bar{1}0 \rangle_c$ and $\langle 11\bar{2} \rangle_c$ directions [36, 55], consistent with the above discussed XRD analysis showing an effective rhombohedral unit cell due to averaging over different monoclinic twin domains oriented along the three degenerate $\langle 1\bar{1}0 \rangle_c$ directions. The XMLD results for LFO on NGO(011)$_o$ and NGO(101)$_o$ are plotted in Figure 9a-b. LFO on NGO is compressively strained along both in-plane directions for either of the two (111)$_{pc}$ facets, but the facets give more or less compressive strain along opposite in-plane axes (see Figure 2). The XMLD signals are weak with a relatively small difference between the $L_{2A/B}$ peaks, consistent with structural domains and in-plane relaxation as seen from the RSM data, indicating the presence of AF domains. Figure 9a shows the data for LFO on NGO(011)$_o$, where the s-polarised signal stays constant with $\theta$, while the p-polarised signal is significantly reduced for $L_{2B}$ in normal incidence. The more pronounced double peak feature at normal incidence indicates a larger Néel vector alignment for normal incidence as compared to either of the grazing incidence measurements. The s-polarised signal indicates a minor component also along the in-plane $[100]_o$. The p-polarised signals, comparing the two different grazing incidence scans at 35° and 145°, suggest a small symmetrical out-of-plane component. However, for grazing incidence scans with $\theta = 15°, 165°$ (not shown), the out-of-plane component vanishes. Thus, it is inferred that the majority of AF domains in the LFO/NGO(011)$_o$ sample are oriented with in-plane Néel vectors. The data is consistent with a resulting AF order oriented along or close to the $[01\bar{1}]_o$ ($[\bar{1}1\bar{2}]_{pc}$) in-plane principal axis. Considering LFO on NGO(101)$_o$ in Figure 9b, the situation is almost identical, the main difference is that the s-polarisation signal is slightly lower at L$_{2B}$, indicating a somewhat larger proportion of domains aligned along the in-plane $[010]_o$. As for LFO on NGO(011)$_o$, the overall picture is that the average over all AF domains are oriented along or close to the $[10\bar{1}]_o$ ($[11\bar{2}]_{pc}$) in-plane principal axis. Hence, for both NGO(011)$_o$ and NGO(101)$_o$, the inferred average LFO spin axis corresponds to the $\langle 11\bar{2}\rangle_{pc}$ direction, the least (1.64 % NGO(011)$_o$) and most (2.00 % NGO(101)$_o$) compressed in-plane axis, respectively. It is noted that when AF domains are present, the inferred spin axis from the XAS data for LFO on NGO does not necessarily reflect the local Néel vector, as the XMLD signal is averaged over many domains due to a beam diameter of roughly 100 μm. The rightmost columns of Figure 9a and b depict both the direction of the averaged magnetic response (blue arrow), as probed by XMLD spectroscopy, and possible local Néel vectors (dashed arrows) compatible with the inferred structural domains (turquoise and yellow).

The XMLD results for LFO on DSO(101)$_o$ are shown in Figure 9c. No significant change in the XAS spectra is seen for any of the three grazing or normal incidence scans. For this situation to occur, the projection of the Néel vector must be the same in all three cases. The p-polarised signal has a relative L$_{2B}$ magnitude above 1.4 in all cases, indicating perpendicular alignment, while the s-polarisation lies below 0.7 with a clear double peak for the high energy tail indicating parallel alignment. This is consistent with a Néel vector aligned with the in-plane *b* axis $[010]_o$, resulting in a monodomain AF signature. This is clearly different from LFO on both facets of NGO, and from LFO on STO(111)$_c$, which results in polydomain AF response. Similar results were obtained for LFO on GSO(101)$_o$, as shown in Figure 9d with an almost identical

XAS response, where the strain conditions and crystal structure are comparable to DSO(101)$_o$. It is noted that the Néel vector direction obtained in these systems, see the rightmost column of Figure 9 for schematics of the Néel vectors, coincides with the axis of large tensile strain, the $[\bar{1}10]_{pc}$ direction that is elongated in-plane.

To further probe the uniaxial magnetic anisotropy of LFO/DSO(101)$_o$, spatially resolved XMLD maps were taken with XPEEM. Figure 10 shows an L$_2$ edge XMLD-XPEEM micrograph where the magnetic domain contrast is uniform, consistent with the XMLD spectroscopy data discussed above. Hence, the data indicate either a sample with domains larger than the field of view (20 µm by 20 µm), or a thin film with macroscopic uniaxial magnetic anisotropy. Literature values for AF domain sizes are typically sub-micron to a few microns of scale, as reported for LFO [33, 36], (La,Sr)FeO$_3$ [58], NiO [59, 60], CuMnAs [16], and Mn$_2$Au [17], substantially smaller than the field of view in the present study. Hence, the XMLD-XPEEM data points towards macroscopic uniaxial anisotropy. It is important to note that 180° AF domains cannot be distinguished in these measurements as they would result in identical response. However, the XRD RSM data is consistent with a crystallographic monodomain state. Together with the structural data, the XAS and spatially homogeneous XPEEM data are thus consistent with an AF monodomain state with a uniaxial Néel vector along the in-plane $[010]_o$ direction, the tensile strained direction.

LFO on DSO(011)$_o$ also exhibits a large magnetic anisotropy, as shown in Figure 9e. Here the s-polarisation signal has increased to around 1.4 and does not change considerably with $\theta$, however the p-polarisation signal is $\theta$-dependent. The low-intensity double peak feature of L$_{2B}$ now occurs distinctively for the p-polarised signal at grazing incidence of $\theta = 35°$, consistent with parallel alignment of the x-ray *E* vector, the $[010]_o$ axis, and the AF spin axis. The p-polarised $L_{2B}$ peaks gradually increases when the x-ray *E* vector becomes less aligned with the $[010]_o$ axis, first fully in-plane for normal incidence and then out-of-plane almost perpendicular at opposite grazing incidence. This is consistent with a Néel vector along the *b* axis $[010]_o$, which is pointing out-of-plane at about 55°. The p-polarised signal goes from below 0.7 to above 1.4, indicating a transition from parallel to perpendicular alignment, while the s-polarised signal stays around or above 1.4 through the series of incidences, indicating little or no alignment with any spin axis. Thus, the data for LFO on DSO(011)$_o$ is also consistent with a uniaxial AF state, with the Néel vector along the $[\bar{1}10]_{pc}$ axis that is here elongated out-of-plane, see the rightmost column of Figure 9e.

All XMLD results are summarised schematically in the rightmost column of Figure 9. LFO deposited on DSO(011)$_o$, DSO(101)$_o$, and GSO(101)$_o$ is found to share the same crystallographic Néel vector orientation. Comparing the two (111)$_{pc}$ facets it is clear that the substrate *b* axis becomes the preferred spin axis in both cases, out-of-plane (red) on (011)$_o$ and in-plane (green) on (101)$_o$ facets. Interestingly, this is the axis where LFO experiences the largest elongation compared to its bulk structure. Having the Néel vector along the longest of the two primary *a*/*b* orthorhombic axes is counterintuitive compared with bulk LFO and the rest of the orthoferrites, where the Néel vector always takes the direction of the shorter *a* axis [25, 27, 61]. However, since the AF superexchange mechanism is sensitive to orbital overlap both energetically and spatially [62, 63], small changes in the Fe-O-Fe bond lengths and buckling angles can significantly alter the AF properties [64]. Albeit perovskites are mainly of ionic character, it is noted that the Néel vector is favoured by the tensile strain direction in intermetallic manganese-based AF systems where the magnetoelastic coefficient is positive [19]. The results thus suggest that the magnetoelastic coefficient in tensile strained LFO thin films is positive, and the large orthorhombic distortion imposed by the scandate substrates clearly forces the magnetic moments along this direction.

## Conclusion

By employing large anisotropic strain, structurally and magnetically monodomain AF thin films are demonstrated. Structurally the LFO thin films, while orthorhombic in bulk, becomes monoclinic or triclinic depending on the orthorhombic substrate of choice. LFO thin films on DSO and GSO exhibit a monodomain structure, and magnetically these films show a macroscopic uniaxial Néel vector along the tensile strained pseudo-orthorhombic $b$ axis, and no magnetic domains are found. The anisotropic strain also effectively lifts the often-observed degeneracy between the orthorhombic $a$ and $b$ axes in thin films, supporting a uniaxial AF state. These findings open for anisotropic strain as a means to design and engineer AF thin films with a specified Néel vector for spin transport-based applications that minimises the effect of AF domain wall spin resistance. It is anticipated that anisotropic strain engineering can be exploited for designing unique system responses that take advantage of anisotropic phenomena, not necessarily limited to insulating antiferromagnets.

## Acknowledgements

The Norwegian Metacenter for Computational Science is acknowledged for providing computational resources, Uninett Sigma 2, Project No. NN9301K and NN9355K. This research used resources at the Advanced Light Source which is supported by the Director, Office of Science, Office of Basic Energy Sciences, of the U.S. Department of Energy under Contract No. DE-AC02-05CH11231. Partial funding was obtained from the Research Council of Norway Grant No. 231290, and the Norwegian Ph.D. Network on Nanotechnology for Microsystems, which is sponsored by the Research Council of Norway, Division for Science, under Contract No. 221860/F60.# References

1. Jungwirth, T., et al., *Antiferromagnetic spintronics.* Nat Nano, 2016. **11**(3): p. 231-241.
2. Baltz, V., et al., *Antiferromagnetic spintronics.* Reviews of Modern Physics, 2018. **90**(1): p. 015005.
3. Gomonay, O., et al., *Antiferromagnetic spin textures and dynamics.* Nature Physics, 2018. **14**(3): p. 213-216.
4. Jungfleisch, M.B., W. Zhang, and A. Hoffmann, *Perspectives of antiferromagnetic spintronics.* Physics Letters A, 2018. **382**(13): p. 865-871.
5. Lebrun, R., et al., *Tunable long-distance spin transport in a crystalline antiferromagnetic iron oxide.* Nature, 2018. **561**(7722): p. 222-225.
6. Wadley, P., et al., *Electrical switching of an antiferromagnet.* Science, 2016. **351**(6273): p. 587.
7. Olejník, K., et al., *Antiferromagnetic CuMnAs multi-level memory cell with microelectronic compatibility.* Nature Communications, 2017. **8**: p. 15434.
8. Bodnar, S.Y., et al., *Writing and reading antiferromagnetic $Mn_2Au$ by Néel spin-orbit torques and large anisotropic magnetoresistance.* Nature Communications, 2018. **9**(1): p. 348.
9. Meinert, M., D. Graulich, and T. Matalla-Wagner, *Electrical Switching of Antiferromagnetic $Mn_2Au$ and the Role of Thermal Activation.* Physical Review Applied, 2018. **9**(6): p. 064040.
10. Olejník, K., et al., *Terahertz electrical writing speed in an antiferromagnetic memory.* Science Advances, 2018. **4**(3): p. eaar3566.
11. Qaiumzadeh, A., et al., *Spin Superfluidity in Biaxial Antiferromagnetic Insulators.* Physical Review Letters, 2017. **118**(13): p. 137201.
12. Qiu, Z., et al., *Spin colossal magnetoresistance in an antiferromagnetic insulator.* Nature Materials, 2018. **17**(7): p. 577-580.
13. Kosub, T., et al., *Purely antiferromagnetic magnetoelectric random access memory.* Nature Communications, 2017. **8**(1): p. 13985.
14. Zheng, J.-H., et al., *Domain-Wall Magnetoresistance in Antiferromagnetic Metals.* arXiv preprint arXiv:2004.07154, 2020.
15. Moriyama, T., et al., *Spin torque control of antiferromagnetic moments in NiO.* Scientific Reports, 2018. **8**(1): p. 14167.
16. Grzybowski, M.J., et al., *Imaging Current-Induced Switching of Antiferromagnetic Domains in CuMnAs.* Physical Review Letters, 2017. **118**(5): p. 057701.
17. Sapozhnik, A.A., et al., *Direct imaging of antiferromagnetic domains in $Mn_2Au$ manipulated by high magnetic fields.* Physical Review B, 2018. **97**(13): p. 134429.
18. Schlom, D.G., et al., *Elastic strain engineering of ferroic oxides.* MRS Bulletin, 2014. **39**(2): p. 118-130.
22

*Table 1:* Complete overview of the crystal structure data found from RSM and DFT for LFO on all substrate facets. Data column describes experimental data from literature or RSM (Ref./Exp.), DFT data relaxed from literature data (Relax.), and DFT data for LFO strained to the respective facets with different LFO supercell initialisations ($a^-a^-c^+$, $a^-a^-c^0$, $a^-a^-a^-$, $a^-a^-c^+$sw). The $a^-a^-c^+$sw initialisation is based on the relaxed DFT cells for each substrate with La/Fe swapped for the respective A/B cations of the substrates. All data are presented on the same format following the orthorhombic orientation of the substrates. The symmetries are deduced from the supercells using FINDSYM [66, 67]. * For STO the LFO structure found from RSM has been given twice, in both rhombohedral and monoclinic symmetry, to facilitate easier comparison of data.

| System | Data | Sym. | Space group | $a$ [Å] | $b$ [Å] | $c$ [Å] | $\alpha$ [°] | $\beta$ [°] | $\gamma$ [°] | $V$ [Å$^3$] | Glazer | $\Delta E$ [meV/f.u.] |
|---|---|---|---|---|---|---|---|---|---|---|---|---|
| Bulk LFO | Ref. [30] | O | 62-3 *Pbnm* | 5.553 | 5.563 | 7.867 | 90 | 90 | 90 | 243.022 | $a^-a^-c^+$ | |
| | Relax. | O | 62-3 *Pbnm* | 5.5595 | 5.5630 | 7.8458 | 90 | 90 | 90 | 242.6483 | $a^-a^-c^+$ | 0.0 |
| DSO(101)$_o$ | Exp. | M | 14-8 *P21/n* | 5.57 | 5.73 | 7.83 | 90 | 90.04 | 90 | 249.52 | | |
| | $a^-a^-c^+$ | M | 14-8 *P21/n* | 5.5271 | 5.6895 | 7.7837 | 90 | 89.8980 | 90 | 244.7690 | $a^-a^-c^+$ | +13.9 |
| | $a^-a^-c^0$ | M | 14-8 *P21/n* | 5.5271 | 5.6895 | 7.7837 | 90 | 89.8985 | 90 | 244.7668 | $a^-a^-c^+$ | +13.9 |
| | $a^-a^-a^-$ | M | 15-3 *I2/a* | 5.4929 | 5.6895 | 7.7781 | 90 | 90.2091 | 90 | 243.0750 | $a^-a^-c^0$ | +25.1 |
| | $a^-a^-c^+$sw | M | 14-8 *P21/n* | 5.4396 | 5.6895 | 7.9057 | 90 | 89.2626 | 90 | 244.6485 | $a^-a^-c^+$ | +26.4 |
| DSO(011)$_o$ | Exp. | M | 14-7 *P21/b* | 5.45 | 5.60 | 7.96 | 90.40 | 90 | 90 | 243.14 | | |
| | $a^-a^-c^+$ | M | 14-7 *P21/b* | 5.3885 | 5.6278 | 7.9316 | 89.8292 | 90 | 90 | 240.5269 | $a^-a^-c^+$ | +34.9 |
| | $a^-a^-c^0$ | M | 14-7 *P21/b* | 5.3885 | 5.6278 | 7.9316 | 89.8290 | 90 | 90 | 240.5277 | $a^-a^-c^+$ | +34.9 |
| | $a^-a^-a^-$ | M | 15-9 *I2/c* | 5.3885 | 5.5720 | 7.9048 | 90.5098 | 90 | 90 | 237.3282 | $a^-a^-c^-$ | +66.8 |
| | $a^-a^-c^+$sw | M | 14-7 *P21/b* | 5.3885 | 5.6831 | 7.8659 | 90.0935 | 90 | 90 | 240.8800 | $a^-a^-c^+$ | +38.4 |
| GSO(101)$_o$ | Exp. | M | 14-8 *P21/n* | 5.57 | 5.75 | 7.83 | 90 | 90.50 | 90 | 250.63 | | |
| | $a^-a^-c^+$ | M | 14-8 *P21/n* | 5.5213 | 5.7203 | 7.7966 | 90 | 90.3416 | 90 | 246.2379 | $a^-a^-c^+$ | +21.7 |
| | $a^-a^-c^0$ | M | 14-8 *P21/n* | 5.5212 | 5.7203 | 7.7966 | 90 | 90.3424 | 90 | 246.2344 | $a^-a^-c^+$ | +21.7 |
| | $a^-a^-a^-$ | M | 15-3 *I2/a* | 5.4819 | 5.7203 | 7.7876 | 90 | 90.7301 | 90 | 244.1817 | $a^-a^-c^{0(-\delta)}$ | +35.7 |
| | $a^-a^-c^+$sw | M | 14-8 *P21/n* | 5.4450 | 5.7203 | 7.9013 | 90 | 89.8046 | 90 | 246.0997 | $a^-a^-c^+$ | +29.4 |
| GSO(011)$_o$ | $a^-a^-c^+$ | M | 14-7 *P21/b* | 5.4315 | 5.6146 | 7.9367 | 90.3159 | 90 | 90 | 242.0286 | $a^-a^-c^+$ | +20.0 |
| | $a^-a^-c^0$ | M | 14-7 *P21/b* | 5.4315 | 5.6146 | 7.9367 | 90.3153 | 90 | 90 | 242.0311 | $a^-a^-c^+$ | +20.0 |
| | $a^-a^-a^-$ | M | 15-9 *I2/c* | 5.4315 | 5.5557 | 7.9079 | 91.0462 | 90 | 90 | 238.5889 | $a^-a^-c^-$ | +47.3 |
| | $a^-a^-c^+$sw | M | 14-7 *P21/b* | 5.4315 | 5.6779 | 7.8613 | 90.6198 | 90 | 90 | 242.4272 | $a^-a^-c^+$ | +25.2 |
| NGO(101)$_o$ | Exp. | T | 2 *P-1* | 5.57 | 5.52 | 7.81 | 89.74 | 88.26 | 89.64 | 239.84 | | |
| | $a^-a^-c^+$ | M | 14-8 *P21/n* | 5.5658 | 5.5342 | 7.7596 | 90 | 88.2552 | 90 | 238.9002 | $a^-a^-c^+$ | +25.7 |
| | $a^-a^-c^0$ | M | 14-8 *P21/n* | 5.5657 | 5.5342 | 7.7596 | 90 | 88.2557 | 90 | 238.8974 | $a^-a^-c^+$ | +25.7 |
| | $a^-a^-a^-$ | M | 15-3 *I2/a* | 5.5529 | 5.5342 | 7.7688 | 90 | 88.2542 | 90 | 238.6269 | $a^-a^-a^-$ | +32.8 |
| | $a^-a^-c^+$sw | M | 14-8 *P21/n* | 5.5405 | 5.5342 | 7.7955 | 90 | 88.0713 | 90 | 238.8908 | $a^-a^-c^+$ | +27.9 |
| NGO(011)$_o$ | Exp. | M | 14-7 *P21/b* | 5.53 | 5.59 | 7.81 | 89.40 | 90 | 90 | 241.67 | | |
| | $a^-a^-c^+$ | M | 14-7 *P21/b* | 5.4078 | 5.6081 | 7.8253 | 88.2094 | 90 | 90 | 237.2014 | $a^-a^-c^{+(-\delta)}$ | +56.4 |
| | $a^-a^-c^0$ | M | 14-7 *P21/b* | 5.4078 | 5.6080 | 7.8252 | 88.2103 | 90 | 90 | 237.1972 | $a^-a^-c^{+(-\delta)}$ | +56.4 |
| | $a^-a^-a^-$ | M | 15-9 *I2/c* | 5.4078 | 5.5928 | 7.8269 | 88.2998 | 90 | 90 | 236.6182 | $a^-a^-c^-$ | +71.2 |
| | $a^-a^-c^+$sw | M | 14-7 *P21/b* | 5.4078 | 5.6446 | 7.7814 | 88.3895 | 90 | 90 | 237.4305 | $a^-a^-c^{+(-\delta)}$ | +58.0 |
| STO(111)$_c$ | Exp.* | R | 167-2 *R-3c* | 5.58 | | | 59.34 | | | 120.88 | | |
| | Exp.* | M | 14-8 *P21/n* | 5.58 | 5.52 | 7.85 | 90 | 89.19 | 90 | 241.77 | | |
| | $a^-a^-c^+$ | M | 14-8 *P21/n* | 5.5739 | 5.5078 | 7.8361 | 90 | 89.0403 | 90 | 240.5332 | $a^-a^-c^+$ | - |
| Bulk DSO | Ref. [38] | O | 62-3 *Pbnm* | 5.4494 | 5.7263 | 7.9132 | 90 | 90 | 90 | 246.9306 | $a^-a^-c^+$ | |
| | Relax. | O | 62-3 *Pbnm* | 5.3885 | 5.6895 | 7.8706 | 90 | 90 | 90 | 241.2944 | $a^-a^-c^+$ | - |
| Bulk GSO | Ref. [38] | O | 62-3 *Pbnm* | 5.4862 | 5.7499 | 7.9345 | 90 | 90 | 90 | 250.2946 | $a^-a^-c^+$ | |
| | Relax. | O | 62-3 *Pbnm* | 5.4315 | 5.7203 | 7.8920 | 90 | 90 | 90 | 245.2026 | $a^-a^-c^+$ | - |
| Bulk NGO | Ref. [39] | O | 62-3 *Pbnm* | 5.4332 | 5.5034 | 7.7155 | 90 | 90 | 90 | 230.7017 | $a^-a^-c^+$ | |
| | Relax. | O | 62-3 *Pbnm* | 5.4078 | 5.5342 | 7.7017 | 90 | 90 | 90 | 230.4914 | $a^-a^-c^+$ | - |
| Bulk STO | Ref. [40] | C | 221 *Pm-3m* | 3.9049 | | | 90 | | | 59.5429 | $a^0a^0a^0$ | |
| | Relax. | C | 221 *Pm-3m* | 3.8946 | | | 90 | | | 59.0743 | $a^0a^0a^0$ | - |

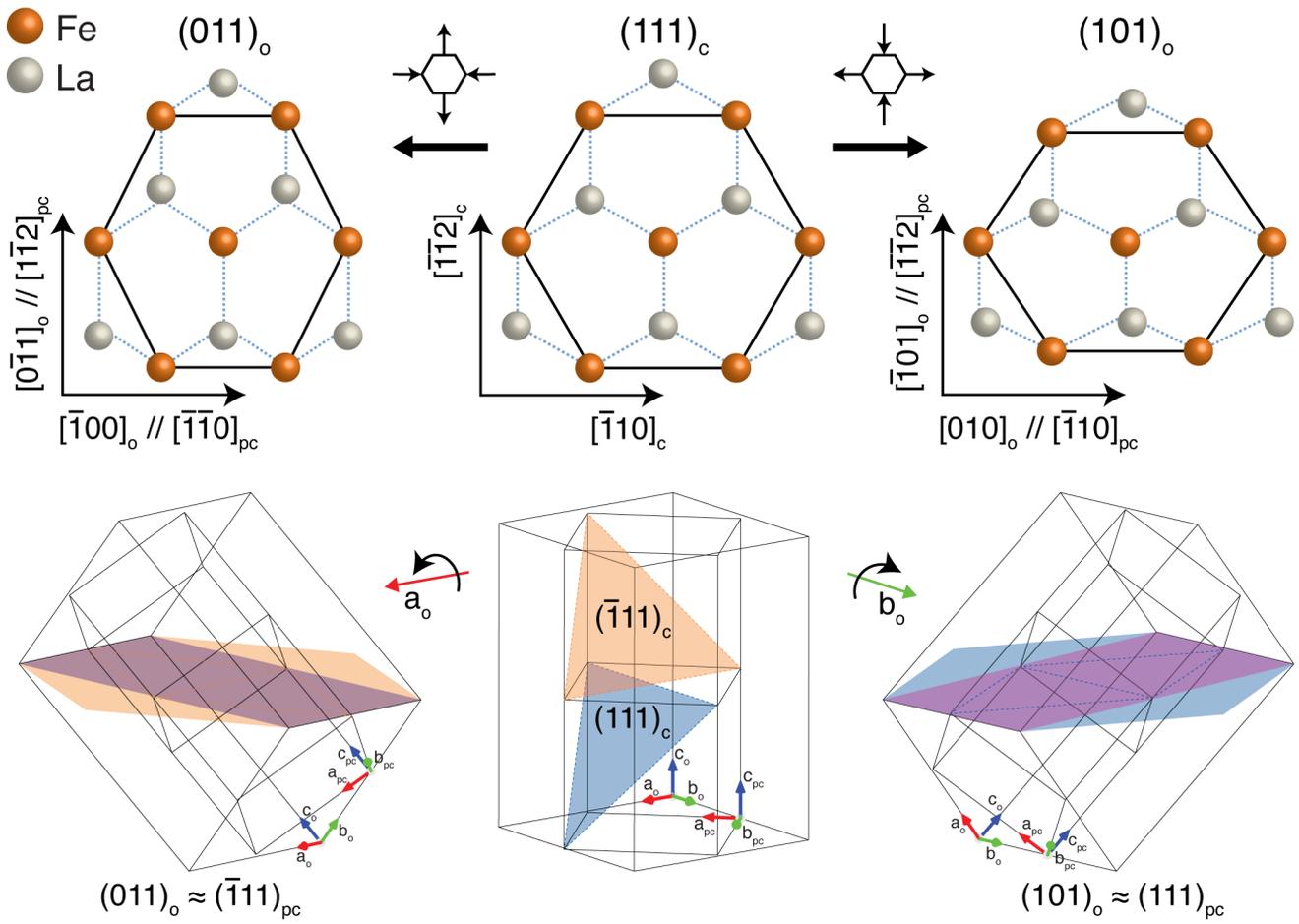

*Figure 1:* **Upper row:** Orthorhombic equivalents of the cubic (111) plane and their effect on strain anisotropy along different crystallographic axes in the (111) pseudo-cubic plane. Upon going from the cubic symmetry to either of the two orthorhombic equivalents, one can note an opposite deformation of the buckled hexagons along the same pseudo-cubic directions. In addition, the orthorhombic symmetry accommodates an antipolar distortion of the A-cations along the $b_o$ axis, which can be noted as successive shifts of La atoms up or down for $(011)_o$ and right or left for $(101)_o$. All distortions have been exaggerated for clarity. **Lower row:** the $(011)_o$ and $(101)_o$ orthorhombic planes (purple and pink shades) along with the respective $(111)_{pc}$ planes (orange and blue shades) shown together with the orthorhombic and cubic unit cells. The orthorhombic unit cell has approximately twice the height and four times the volume of a cubic unit cell.

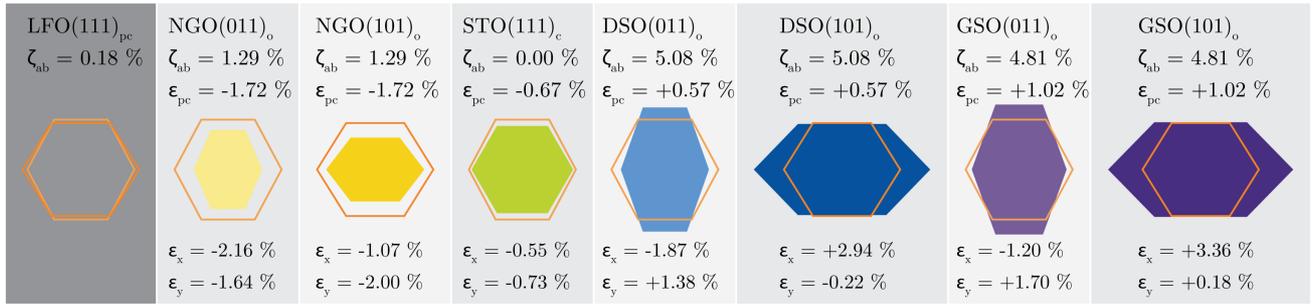

*Figure 2:* Overview of strain conditions for LFO (011)$_o$ (light orange outline) and (101)$_o$ (orange outline) on all substrate facets (filled colors). The (011)$_o$ facet of LFO has been used for STO(111)$_c$. The figures are proportionally exaggerated in order to clearly convey the differences. Note that the level of orthorhombic distortion, denoted by $\zeta_{ab}$, is small in LFO, larger in NGO and largest in the scandates.

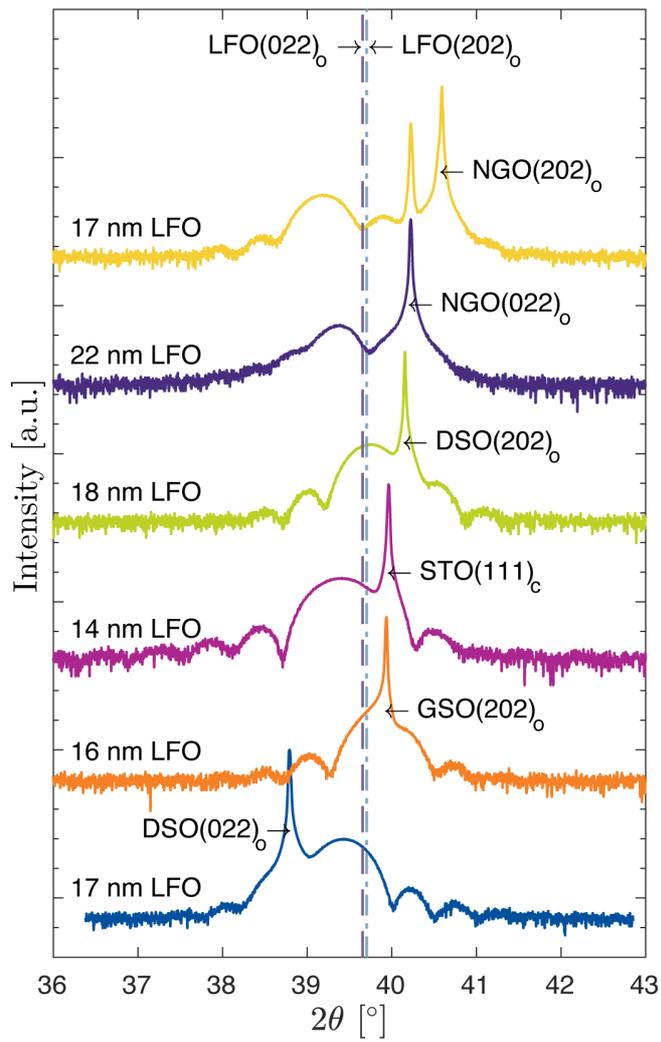

*Figure 3:* Symmetric $\theta/2\theta$ scans for all samples offset for clarity. Each sample is labelled above the scan data on the left side, and the two vertical dashed lines represent the $(111)_{pc}$ planes of bulk LFO as labelled by the arrowed text on top. A straightforward relationship between substrate and film peak positions cannot be expected for $(111)_{pc}$ oriented interfaces, as the strain can be accommodated along directions other than the surface normal. Hence, there is not a monotonic shift of the LFO peak position when going from substrates with large $d_{(111)pc}$ for $DSO(011)_o$, towards decreasing $d_{(111)pc}$ for $NGO(101)_o$.

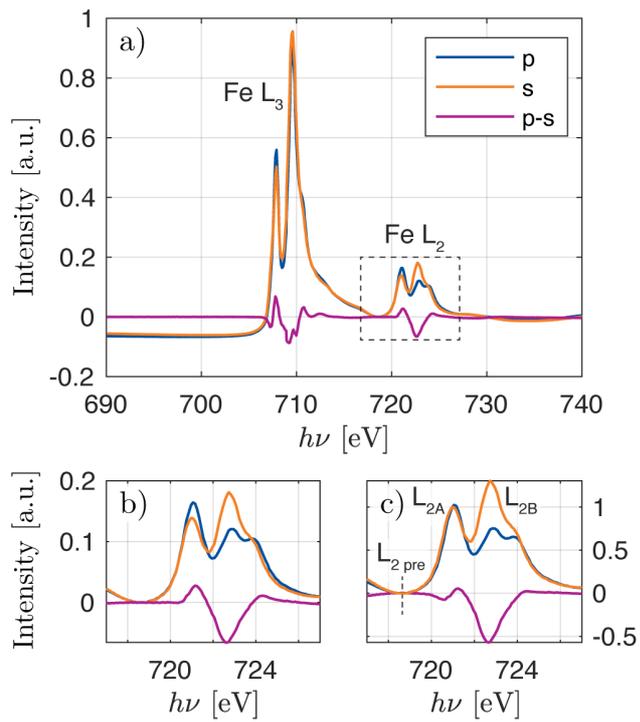

*Figure 4:* (a) X-ray absorption spectra for s and p polarized x-rays taken at the Fe $L_{2/3}$ edges. The XMLD effect is represented by the difference signal p-s between polarisations. (b) Zoomed in view of the $L_2$ multiplet with normalised data, and (c) zoomed in view of the $L_2$ multiplet after renormalisation with the $L_2$ pre-edge set to zero and the $L_{2A}$ peaks set to one.

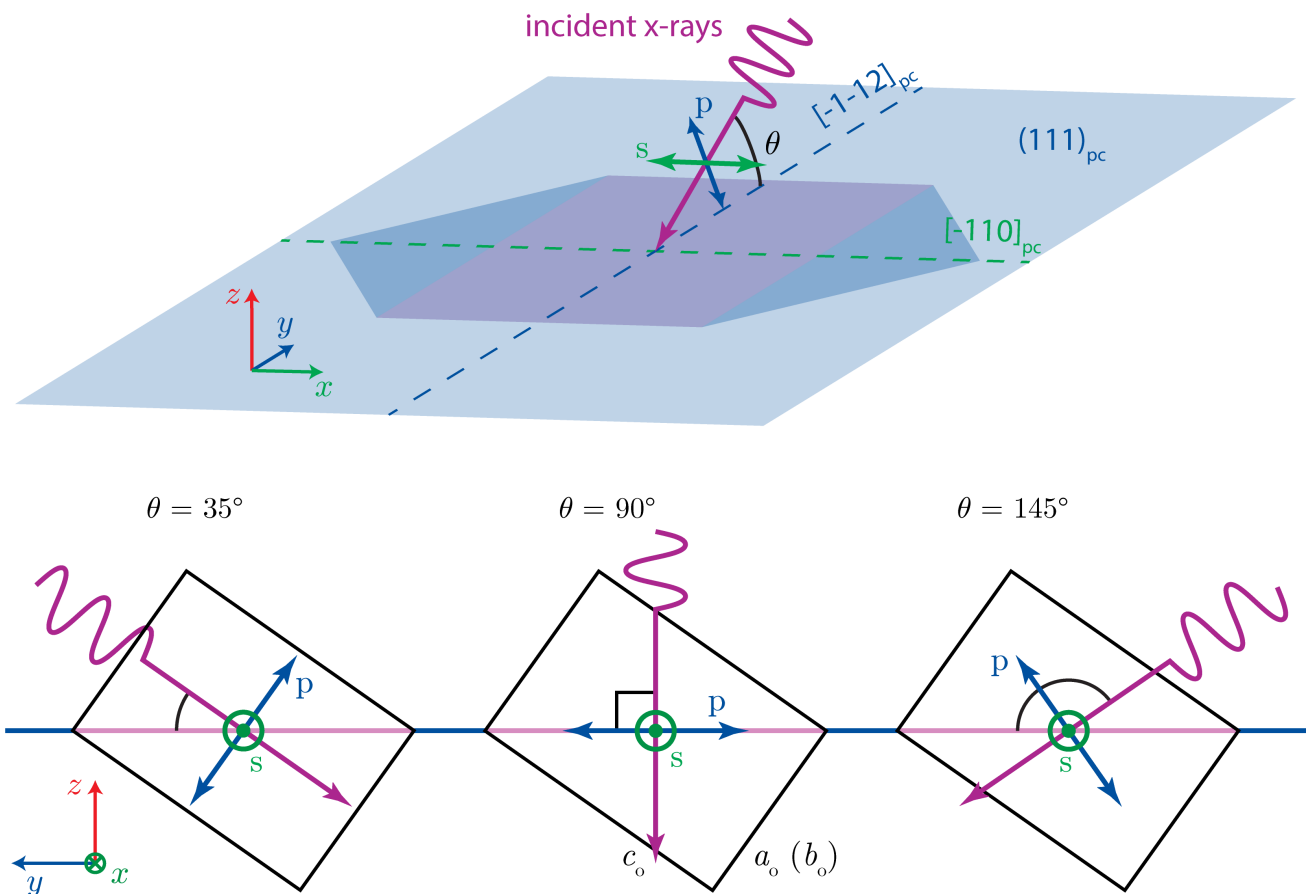

*Figure 5:* **Top:** XAS geometry for XMLD measurements. Incoming x-rays are linearly polarised with either s- or p-polarisation. Incidence angle varies with $\theta \in [15°, 165°]$ and the azimuthal direction is fixed to have the plane of incidence parallel to the $(1\bar{1}0)_{pc}$ symmetric plane and the $[11\bar{2}]_{pc}$ principal in-plane axis. The buckled hexagon $(111)_{pc}$ surface and the orthorhombic equivalent plane are indicated in blue and pink shades, respectively. The figures are made with reference to the $(101)_o$ orientation of the orthorhombic unit cell, and the coordinate system is defined as $\hat{x} \parallel [\bar{1}10]_{pc}$, $\hat{y} \parallel [\bar{1}\bar{1}2]_{pc}$, $\hat{z} \parallel [111]_{pc}$. **Bottom:** The three most used incidence angles are drawn as seen along the $b_o$ ($a_o$) axis. Upon changing the incidence angle, p-polarisation probes along different directions in the plane of incidence, while s-polarisation stays fixed along $b_o$ ($a_o$).

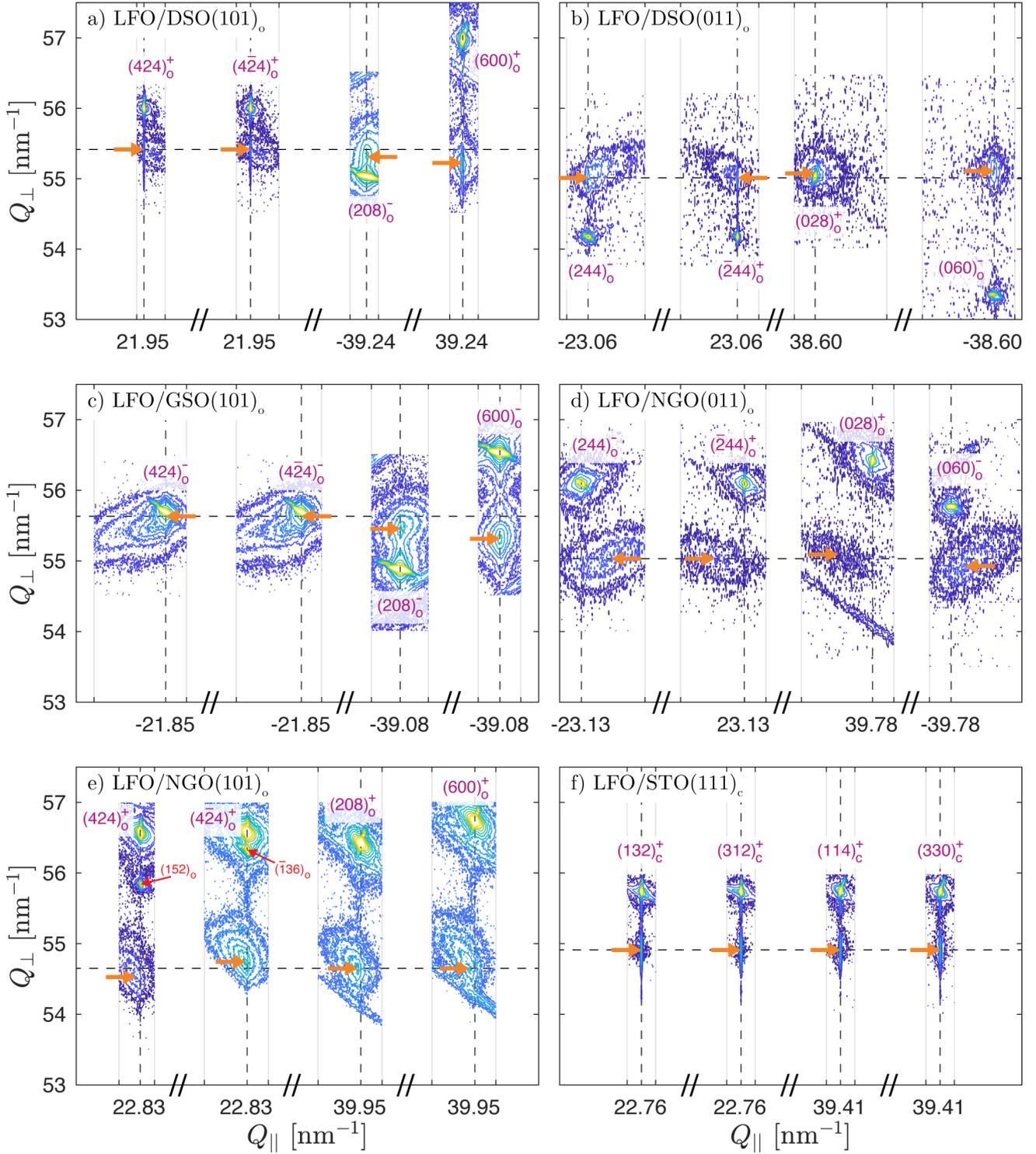

*Figure 6:* RSM data for all samples. LFO on DSO(101)$_o$ (a), DSO(011)$_o$ (b), GSO(101)$_o$ (c), NGO(011)$_o$ (d), NGO(101)$_o$ (e), STO(111) (f). Axes are 1:1, and all plots are to scale with one another for direct comparison. X-axis values and dashed vertical lines denote the substrate $Q_\parallel$ for each reflection. Grey vertical lines show the in-plane limits of the RSM. The horizontal dashed lines are guides to the eye for comparison of the film reflection values $Q_\perp$. The film peaks are found at the orange arrowheads. The streaks that are seen around $Q_\perp \approx 54$ and $Q_\perp \approx 56.5$ in some of the high $Q_\parallel$ scans have been verified to originate from the XRD system itself – they have nothing to do with the samples.

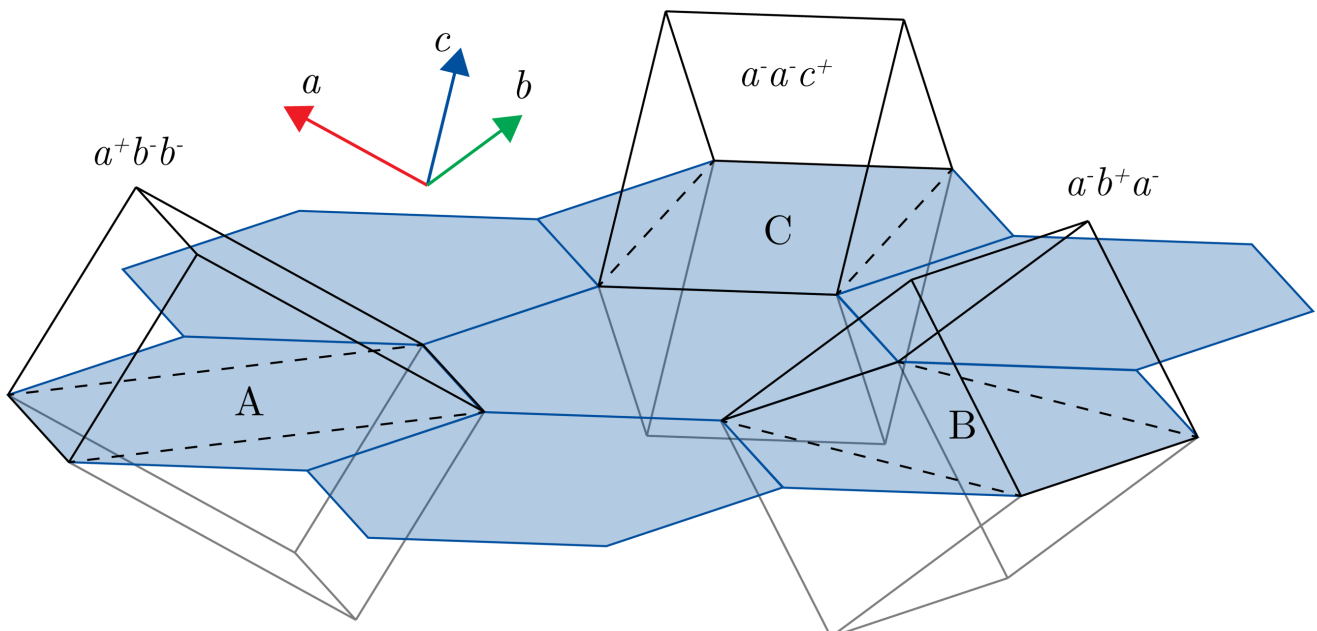

*Figure 7:* The three structural domains an orthorhombic film material can take on a pseudo-cubic (111)-oriented surface. Arrows indicate the pseudo-cubic unit vectors, and the Glazer tilt system relative to these are stated for each structural domain. The domains are denoted A, B and C corresponding to the orientation of the long axis with in-phase octahedral rotations. If crystal twinning occurs, all three domains may have twins with a 90° rotation around the respective long axes ($a_o$/$b_o$ interchanged) to a total of six structural domains A, A', B, B', C, C'.

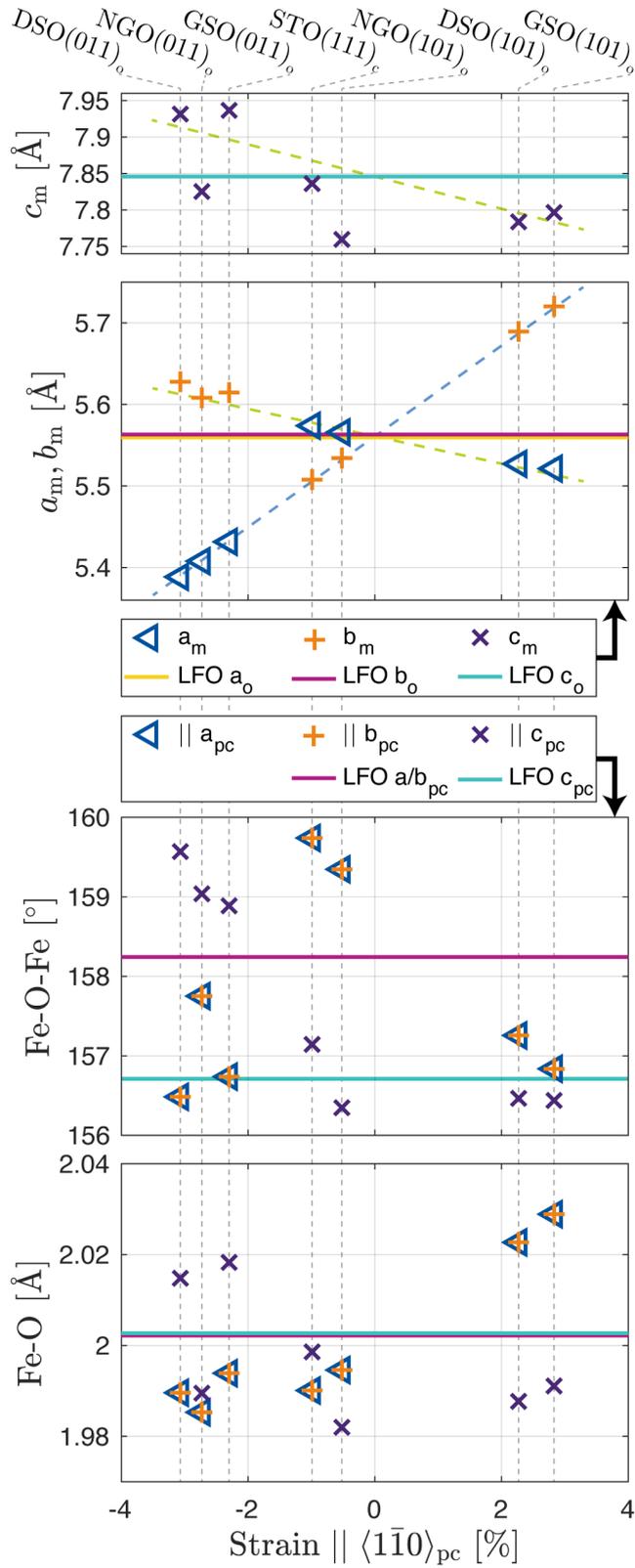

*Figure 8:* Visualisation of the DFT data for LFO on all $(111)_{pc}$ facets plotted with respect to the in-plane strain along the x-axis ($a_o$ for $(011)_o$ and $b_o$ for $(101)_o$ facets). The monoclinic unit cell parameters are found in the upper two panels. The lower two panels show the Fe-O-Fe buckling angles and Fe-O bond lengths along the respective pseudo-cubic axes. Solid lines denote the bulk equilibrium values for the orthorhombic LFO found by DFT.

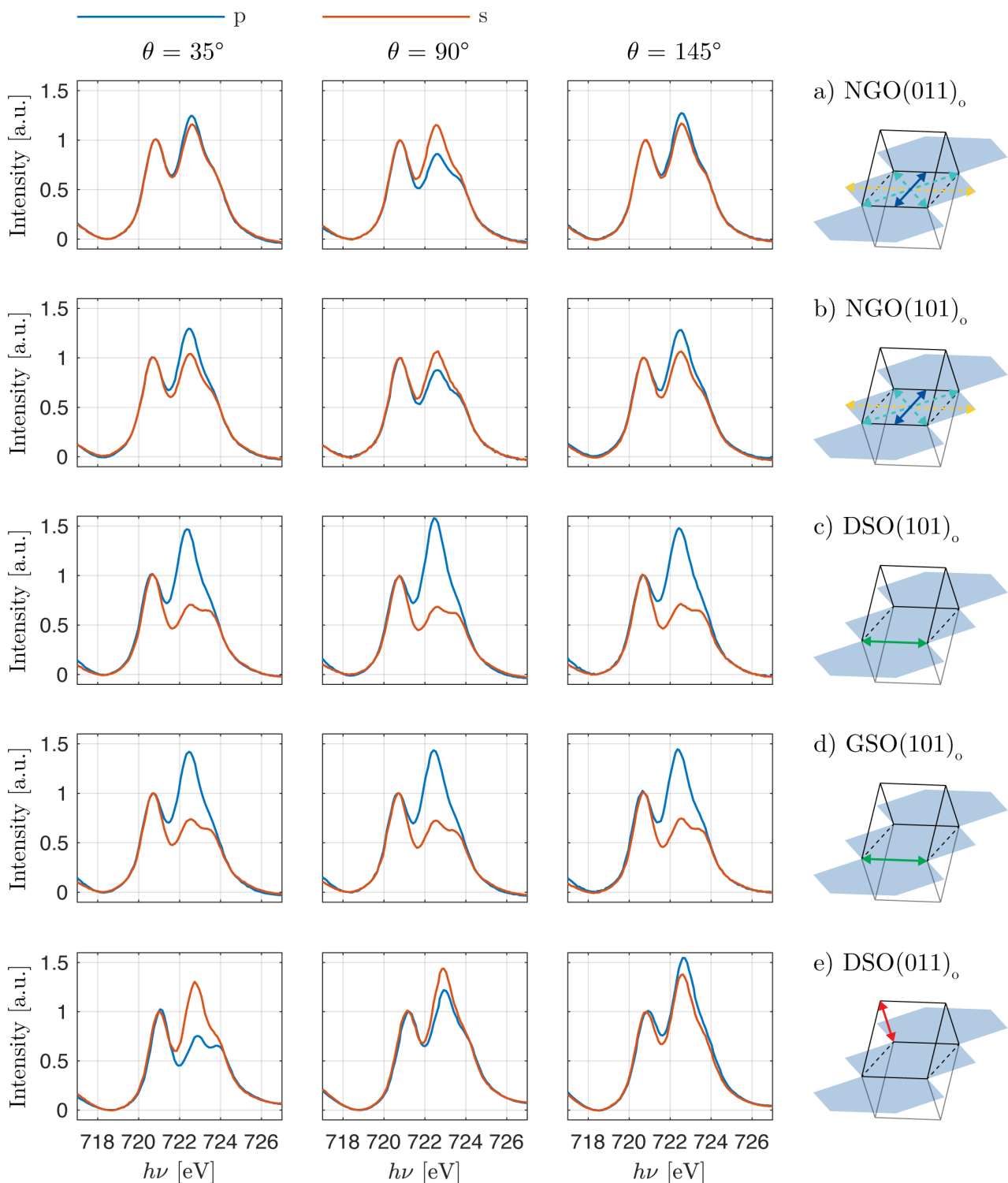

*Figure 9:* Fe L$_2$ XAS spectra for p- and s-polarised x-rays normalised at the L$_{2A}$ and L$_2$ pre-edge for LFO on all facets (rows); NGO(011)$_o$ (a), NGO(101)$_o$ (b), DSO(101)$_o$ (c), GSO(101)$_o$ (d), DSO(011)$_o$ (e). The four columns show from left to right data for increasing x-ray incidence angle and the resulting Néel vector orientation relative to the substrate facets in the rightmost column. Solid arrows indicate the resulting spin axis from the XMLD data, and dashed arrows indicate possible local Néel vectors in the cases where a monodomain signature was not found.

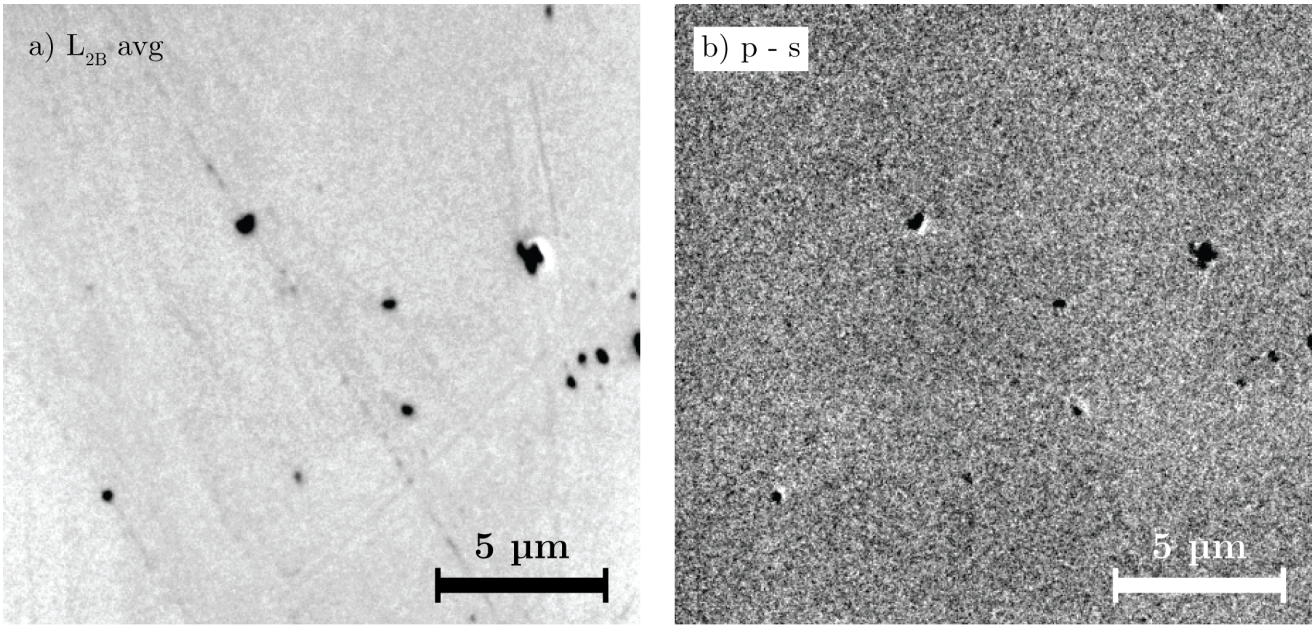

*Figure 10:* XMLD-PEEM micrographs of the LFO/DSO(101)$_o$ sample showing no sign of AF domain contrast, indicating a magnetic monodomain over the entire field of view (18.4 x 18.4 µm²). Images were taken with both s- and p-polarisation, with the average of the L$_{2B}$ peak signal seen in a), and the XMLD difference image in b). The black dots and line features seen particularly in a) are contaminants and scratches due to handling between several different experiments.